\documentclass
[aps,prd,floats,balancelastpage,prd,amssymb,floatfix,nofootinbib,amsmath]{revtex4}

\usepackage{epsfig}
\usepackage{epstopdf}

\usepackage{float}
\usepackage[usenames]{color}
\usepackage{color,epsfig,amsmath,amssymb,fancyhdr}   
\usepackage[french]{babel}
\usepackage{ulem}   

\begin{document}

\title{Reply to comment on  ``Irregularity in gamma ray source spectra as a signature of axionlike particles"}

\author{Pierre Brun}
\email{pierre.brun@cea.fr}
\affiliation{CEA, Irfu, Centre de Saclay, F-91191
Gif-sur-Yvette | France}

\author{Denis Wouters}
 \email{denis.wouters@cea.fr}
\affiliation{CEA, Irfu, Centre de Saclay, F-91191
Gif-sur-Yvette | France}

\begin{abstract}
G. Galanti and M .Roncadelli recently made public some comments on the article by D. Wouters and P. Brun about irregularities induced by photon mixing to axion-like particles in astrophysical media [Phys. Rev. D {\bf 86}, 043005 (2012)]. They claim in particular to have found some mistakes in the article. This note is a response to their comments, we refute their arguments and show that the results presented in the article are correct. It turns out most of the misunderstandings come from the definition of the beam initial state, some clarifications about which are given here.
\end{abstract}

\pacs{14.80.Va, 98.70.Vc}

\maketitle

\subsection*{Equation of propagation}
As shown in \cite{1} and used later e.g. in~\cite{2}, the propagation of a beam of a mixed state of photon and axion-like particle (ALP) along the $z$ axis can be described by

\begin{equation}
\left ( \partial^2_z +E^2 + 2E\mathcal{M} \right )\;\psi \;=\;0\;\;,
\end{equation}
where $\mathcal{M}$ is the mixing matrix shown in Eq. 3 of~\cite{4}:
\begin{equation}
\mathcal{M}\;\;=\;\;\left(\begin{array}{ccc} -\omega_p^2/2E-i\Delta_{\mathrm{abs}} & 0 & \Delta_\mathrm{B}\cos\phi \\ 0 & -\omega_p^2/2E-i\Delta_{\mathrm{abs}} & \Delta_\mathrm{B}\sin\phi \\ \Delta_\mathrm{B}\cos\phi & \Delta_\mathrm{B}\sin\phi\ & \Delta_\mathrm{a} \end{array}\right)\;\;,
\end{equation}
This propagation equation can be linearized as in \cite{1}, leading to
\begin{equation}
\left ( E+k \right )\left ( E \pm i \partial_z \right ) \psi + 2E\mathcal{M}\psi\;=\;0\;\;.
\end{equation}
The refractive index being close to 1, one can write $E+k=2E$, which leads to 
\begin{equation}
\left(E \pm i\partial_z + \mathcal{M} \right )
\left(\begin{array}{c} A_1 \\ A_2 \\ a \end{array} \right) = 0 \;\;,
\end{equation}
The different coefficients are the amplitudes for the different polarization states of the photon ($A_i$) and $a$ is the amplitude for the ALP state. The choice of the sign of the derivative is a mere convention regarding the orientation of the axes. This choice has no consequence on the physical quantities that are subsequently computed.

\subsection*{Single domain conversion probability}

In the simple case of a single magnetic domain, the authors of~\cite{3} claim that there is a factor of 2 missing in Eq.~4 of~\cite{4}, compared to their Eq.~6. Whether this factor of 1/2 appears or not actually depends on the photon initial state. To be more precise, their Eq.~6 is valid if the photon beam is taken as fully polarized (electric field) parallel to the transverse component of the external magnetic field. On the contrary, if the beam is assumed to be unpolarized, as it is the case for Fig. 1 of~\cite{4} (see caption), the correct equation is indeed 1/2 of their Eq.~6. 

This can be understood with the formalism of the matrix density  that the authors of~\cite{3} introduce in their comment. Here, we are in the case of an homogeneous magnetic field, so we can define our basis with propagation of the system along the z axis and transverse magnetic field parallel to the y axis. The mixing matrix $\mathcal{M}$ then writes, neglecting supplementary terms $\omega_{\rm pl}$ and $\Delta_{\rm abs}$:
\begin{equation}
\mathcal{M}\;\;=\;\;\left(\begin{array}{ccc} 0 & 0 & 0 \\ 0 & 0 & \Delta_\mathrm{B} \\ 0 & \Delta_\mathrm{B} & \Delta_\mathrm{a} \end{array}\right)\;\;.
\end{equation}
The density matrix $\rho$ after propagation through the magnetic domain can be computed using Eq. 9 of~\cite{3}:
\begin{equation}
\rho = \mathcal{U}\rho_{\rm initial}\mathcal{U}^\dagger\;\;,
\end{equation}
where $\rho_{\rm initial}$ is the initial density matrix and $\mathcal{U}$ is the transfer matrix, that is obtained once the equations of motion are solved.
An initial state fully polarized along the transverse magnetic field direction, 
\begin{equation}
\rho_{\rm initial} = \left(\begin{array}{ccc} 0 & 0 & 0 \\ 0 & 1 & 0\\ 0 & 0 & 0 \end{array} \right)\;\;\text{leads to} \;\; 
P_{\gamma\rightarrow a} \;=\; \frac{4 \Delta_\mathrm{B}^2}{\Delta_{\mathrm{osc}}^2}\sin^2\frac{\Delta_{\mathrm{osc}}z}{2}\;\;,
\end{equation}
and it can then be deduced that for an unpolarized initial state, 
\begin{equation}
\rho_{\rm initial} = \frac{1}{2}\left(\begin{array}{ccc} 1 & 0 & 0\\ 0 & 1 & 0\\ 0 & 0 & 0 \end{array} \right)\;\;\text{leads to} \;\; 
P_{\gamma\rightarrow a} \;=\; \frac{2 \Delta_\mathrm{B}^2}{\Delta_{\mathrm{osc}}^2}\sin^2\frac{\Delta_{\mathrm{osc}}z}{2}\;\;,
\end{equation}
since only the lower-right 2$\times$2 block is included in the mixing. Moreover, if the initial beam is assumed to be fully polarized along the first direction,
\begin{equation}
\rho_{\rm initial} = \left(\begin{array}{ccc} 1 & 0 & 0\\ 0 & 0 & 0\\ 0 & 0 & 0 \end{array} \right)\;\;\text{leads to} \;\; 
P_{\gamma\rightarrow a} \;=\; 0\;\;,
\end{equation}
which is expected since only the parallel component of the polarization couples.
\\ \ \\
In~\cite{4}, the second case is considered. Eq. 4 of~\cite{4} is thus correct.

\subsection*{Cosmological effects}

In realistic conditions, several magnetic domains with different orientations have to be considered. In their comment paper, the authors of~\cite{3} state that in~\cite{4} ``WB do not take cosmological effect into account and assume that all domains have the same size $s$". This is not correct.  We recall here the sentence from~\cite{4} : ``To account for redshifting, a flat $\rm \Lambda CDM$ Universe with $(\Omega_{\rm \Lambda},\,\Omega_{\rm m})=(0.73,\,0.27)$ and $H_0=71\,\rm km/s/Mpc$ is assumed". The physical sizes of the domains as well as the energy, the magnetic field strengths and the plasma density are of course corrected by the relevant powers of $(1+z)$. It was assumed that such an obvious effect was to be implicitly understood by the reader as taken into account.

\subsection*{Absorption on the EBL}

One area of concern expressed in~\cite{3} is the way the absorption on the EBL is taken into account. First we need to stress that contrarily to what the authors of~\cite{3} seem to mean in their comment, the EBL level and spectrum is not everywhere the same. It has a non trivial dependence on the redshift (see for instance Fig. 4 of~\cite{fran}). This means that the mean free path of the photons will not be the same in every magnetic domain along the line of sight. In this regard, the sentence from~\cite{4}: ``\dots ~assuming a propagation over a domain of size $s$ within which the opacity is homogeneous", expresses that we make the (justified) assumption that the magnetic domains are small enough so that the density of the EBL is homogeneous in the domain. It is important to check that $\Delta_\mathrm{abs}$ does not depend on $z$ since otherwise the equations of motion cannot be solved as they are usually. 

The absorption term is introduced by the unitarity-breaking terms in the upper-left diagonal coefficients $-i\Delta_{\rm abs} = -i\tau/2s$. Note that the sign of this term is conditioned by the axis convention mentioned in the first section, and has to be consistent with the used propagation equation. Putting out the first line of Eq. 2 of~\cite{4} and unplugging the photon/ALP mixing by considering $\Delta_{\rm B}=0$ one gets for the first component of the photon:
\begin{equation}
-i\partial_z A_1 + E A_1 -i\Delta_{\rm abs} = 0\;\;.
\end{equation}
The solution for $A_1$ is 
\begin{equation}
A_1=A_{1,(0)} e^{ -iEz} e^{  -\Delta_{\rm abs}z}\;\;.
\end{equation}
In the end the EBL absorption over a domain of size $s$ is such that the flux is reduced as $\phi=\phi_0\times e^{-\tau}$, $\tau$ being the optical depth in the domain. The absorption is then evaluated with
\begin{equation}
\left | \langle A_1 | A_1 \rangle \right |_{z=s}^2\;=\;A_{1,(0)}^2 e^{-2\Delta_{\rm abs} s}\;\;.
 \end{equation}
Then the correct absorption is retrieved by fixing $\Delta_{\rm abs}= \tau /2 s$, as correctly done in~\cite{4}.

\subsection*{Amplitudes of the irregularities and initial state}

The authors of~\cite{3} claim there is a mistake in Fig. 2 of~\cite{4} as for an unpolarized beam $P_{\gamma\rightarrow\gamma}$ is bound to be greater than 1/2. This later statement is true in the case of an unpolarized beam. Unfortunately the authors of~\cite{3} assumed Fig. 2 was produced for an unpolarized beam whereas it is stated nowhere in the paper that it is the case. This point might have required some clarification in the first place but actually for this figure, the initial state that was used is $\psi_{\rm initial}=\left\{ 1;0;0\right \}$. So the beam is indeed polarized and in that case it is correct that the probability $P_{\gamma\rightarrow\gamma}$ can drop below 1/2.

In the comment paper, the question is raised whether the system is described with amplitudes or by the mean of the density operator. We argue here that is does not matter. Owing to the fortunate consistency of quantum mechanics, both approaches lead to the same results. First, one can choose to describe the system using single photons. In that case, the initial state has to include a random angle corresponding to a direction along which a superposition of two polarization states is considered. This can be implemented for example by using the following initial state:
\begin{equation}
\psi_{\rm initial}\;=\;\left(\begin{array}{c} \cos \theta_r \\ \sin \theta_r \\ 0 \end{array} \right)\;\;,
\end{equation}
where $\theta_r$ is a random angle uniformly distributed in $[0,2\pi[$. A second approach consists in acknowledging the lack of information about the initial state and use the density matrix. The underlying idea of using the density matrix is to define directly probabilities as the initial state instead of amplitudes. Of course this can only be understood in the case of a  very large number of photons, to describe a statistical ensemble of photons about which the observer has only incomplete knowledge. One can easily check that in the large photon number limit, the single photon approach gives the same $P_{\gamma\rightarrow\gamma}$ pattern as the one obtained with the density matrix. This is shown on Fig. 1 below, where for simplicity we assumed $\Delta_{\rm abs}=0$. In both cases, the initial state is unpolarized, which is why $P_{\gamma\rightarrow\gamma}\geq 1/2$. The top panel is obtained with the density matrix formalism, and the middle panel is obtained with random initial directions for 10,000 photons. Both figures are obtained with the same realization of a galaxy cluster type of magnetic field, with a 10 kpc coherence length, a RMS intensity of $1\;\rm\mu G$, and ALPs of mass $m= 30\;\rm neV$ and coupling $g=5\times 10^{-11}\;\rm GeV^{-1}$. 
\begin{figure}[b]
\centering
\includegraphics[width=.7\textwidth]{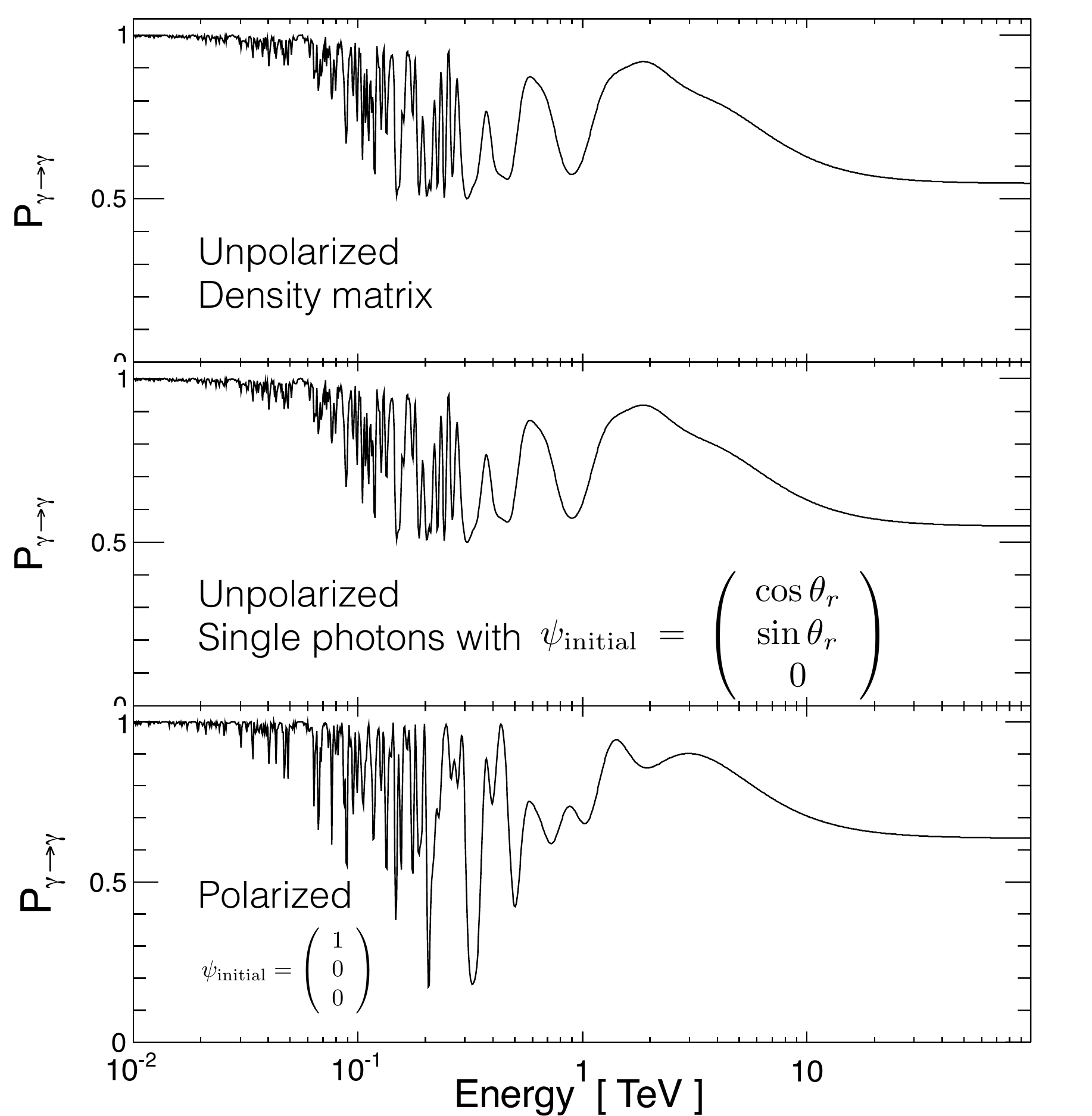}
\caption{Irregularity patterns obtained with different formalisms and assumptions for the initial state.}
\end{figure}
On Fig. 1 it appears clearly that the irregularity pattern that is predicted is the same whatever the used formalism, as long as the large photon number limit is attained in the case of the description with amplitudes. Note also that contrary to what the authors of~\cite{3} seem to fear, none of the wrong equations of~\cite{2} were used to obtain the results from~\cite{4}.

\subsection*{Observability of the irregular pattern}

In Fig. 1 we also displayed the irregularity pattern that would be obtained in the same conditions in the case of an initial beam of photons polarized along a fixed direction. One can see that the irregularity pattern is slightly different, in particular with some peaks appearing deeper. This should answer the concerns of the authors of~\cite{3} who seem to think that the irregularity level and its observability depend a lot on the assumption made about the initial state.

Concerning the observability, the smearing of the irregularities by the instrument and their sensitivity, these aspects are correctly addressed in~\cite{4}. Subsequent papers where actual constraints are obtained out of this novel method, assuming an unpolarized initial photon beam, show that the irregularity effect is within the reach of running experiments.

\acknowledgements{
We would like to thank Philippe Brax for reading this note.
}


\begin{thebibliography}{0}
\bibitem{1} G. Raffelt, L. Stodolsky, Phys. Rev. D {\bf 37}, 1237 (1988)
\bibitem{2} A. De Angelis, G. Galanti, M. Roncadelli, Phys. Rev. D {\bf 84}, 105030 (2011)
\bibitem{3} G. Galanti, M. Roncadelli, arXiv:1305.2114 (2013)
\bibitem{4} D. Wouters, P. Brun, Phys. Rev. D {\bf 86}, 043005 (2012)
\bibitem{fran} A. Franceschini, G. Rodighiero, M. Vaccari, \ 2008, Astron. Astrophys., {\bf 487}, 837 
\end{thebibliography}
\end{document}